\documentclass[a4paper,12pt]{article}
\pdfoutput=1 
\usepackage{epsfig,graphicx,amsmath,amssymb}
\usepackage{chngcntr}
\counterwithin{figure}{subsection}
\counterwithin{table}{subsection}
\usepackage{url}
\usepackage{hyperref}
\newcommand{\ee}{e^{+} e^{-}}

\newcommand{\epm}{e^+e^-}

\def \ee {e^+e^-}

\begin{document}

\thispagestyle{empty}

$\phantom{.}$

\hfill

\begin{center}
{\Large {\bf 14th Meeting of the Working Group on Rad. Corrections and MC Generators for Low Energies} \\
\vspace{0.75cm}}

\vspace{1cm}

{\large September 13, 2013 in Frascati, Italy}

\vspace{2cm}

{\it Editors:}
S.~E.~M\"uller (Dresden) and G.~Venanzoni (Frascati)

\vspace{2.5cm}

ABSTRACT

\end{center}

\vspace{0.3cm}

\noindent
The mini-proceedings of the 14th Meeting of the ``Working Group on Rad. Corrections and MC Generators for Low Energies'' held in Frascati on September 13, 2013, as a satellite meeting of the PHIPSI13~\footnote{\url{http://www.roma1.infn.it/phipsi13/}} conference in Rome are presented. These meetings, started in 2006, have as aim to bring together experimentalists and  theorists working in the fields of meson transition form factors, hadronic contributions to $(g-2)_{\mu}$ and the effective fine structure constant, and development of MonteCarlo generators and Radiative Corrections for precision $\ee$ and $\tau$ physics.

\medskip\noindent
The web page of the meeting, which contains all talks, can be found at
\begin{center}
\url{https://agenda.infn.it/conferenceDisplay.py?confId=6618}
\end{center}

\vspace{0.5cm}


\newpage

{$\phantom{=}$}

\vspace{0.5cm}

\tableofcontents

\newpage

\section{Introduction}

\addtocontents{toc}{\hspace{1cm}{\sl H.~Czy\.z and G.~Venanzoni}\par}
\label{sec:Intro}
\vspace{5mm}

\noindent
H.~Czy\.z$^1$ and G.~Venanzoni$^2$

\vspace{5mm}

\noindent
$^1$ Institute of Physics, University of Silesia, 40007 Katowice, Poland\\
$^2$ Laboratori Nazionali di Frascati dell'INFN, 00044 Frascati, Italy\\
\vspace{3mm}

The importance of continuous and close collaboration between the experimental
and theoretical groups is crucial in the quest for
precision in hadronic physics.
This is the reason why the  
Working Group on ``Radiative Corrections and Monte Carlo Generators for Low Energies'' (Radio MonteCarLow)  was formed a few years ago bringing together experts (theorists and experimentalists) working in the field of low-energy $\epm$ physics and partly also the $\tau$ community.
Its main motivation  was to understand the status and the precision of the Monte Carlo generators used to analyse the hadronic cross section measurements
obtained as well with energy scans as with radiative return, to determine luminosities, and whatever possible to perform tuned comparisons, {\it i.e.}
comparisons of MC generators with a common set of input parameters and experimental cuts. This  main effort was summarized in a report published in 2010~\cite{Actis:2010gg}.
During the years the WG structure has been enriched of more physics items 
and now it includes seven subgroups: Luminosity, R-measurement, ISR, 
Hadronic VP incl. $g-2$ and $\Delta \alpha$, gamma-gamma physics, FSR models, tau. 

During the workshop the last achievements of each subgroups have been presented. 
A particular emphasis has been put to the recent evaluations of the Leading order and Light-by-Light hadronic contributions to the $g-2$ of the muon.
Finally the status of MC generators for R-measurement with energy scan, ISR, and tau decays has been discussed.

\medskip\noindent All the information on the WG can be found at the web page:
\begin{center}
\url{http://www.lnf.infn.it/wg/sighad/} 
\end{center}

\newpage

\section{Short summaries of the talks}

\subsection{Generic MC Generator for $e^+e^- \to$ hadrons at $\sqrt{s} < 2$ GeV}
\addtocontents{toc}{\hspace{2cm}{\sl S.~Eidelman, A.~Korobov}\par}
\label{sec:Eidel}
\vspace{5mm}

S.~Eidelman, A.~Korobov 

\vspace{5mm}

\noindent
      Budker Institute of Nuclear Physics SB RAS and  \\
Novosibirsk State University, Novosibirsk, Russia \\ 
\vspace{3mm}

There is a need for a generic MC generator
approximately reproducing a real picture of
$e^+e^- \to$ hadrons below 2 GeV. 
Such generators exist for higher energy ranges: PYTHIA~\cite{pythia}, 
LUARLW~\cite{luarlw}, $\ldots$ 
based on a complicated scheme of quark and gluon hadronization and
provide events of $e^+e^- \to q\bar{q}$, $q=u,d,s,c,b$.
These generators are used for simulation of various processes studied and 
background estimation.
However, at low energy one can't create a generator based on first 
principles, so existing data on cross sections 
should be used.
We have created the first version of the MC generator  
helping in preselection of main backgrounds in analysis of CMD-3 data
using our database of various cross section measurements. Currently about 
30 various final states are taken into account.
If data are not available, we use isospin relations. 

The algorithm is the following. Energy dependence of the cross section 
for each exclusive final state is approximated
by a physically motivated  analytic function $f_i(s)$. 
At each energy $\sqrt{s}$ event generation includes a calculation of the 
total cross section $\sigma_{\rm tot}(s)=\Sigma{f_i(s)}$; sampling of
a random number specifying the final state to sample; sampling of 
an event of the specific process  based on
the corresponding dynamics.

A list of things to do includes an increase of the number of processes,
using more data, improving isospin relations, taking into account dynamics
(from phase space to real matrix elements).
It is also planned to include approximately an ISR photon.

This work is supported by the Ministry of Education and Science of the
Russian Federation, the RFBR grants  11-02-00112, 11-02-00558 and 
the DFG grant HA 1457/7-2.

\newpage

\subsection{Luminosity measurement with CMD-3 detector at the VEPP-2000 $e^+e^-$ collider}
\addtocontents{toc}{\hspace{2cm}{\sl G.~V.~Fedotovich}\par}
\label{sec:Fedo}
\vspace{5mm}

G.~V.~Fedotovich

\vspace{5mm}

\noindent
     Budker Institute of Nuclear Physics SB RAS and  \\
Novosibirsk State University, Novosibirsk, Russia \\ 

\vspace{3mm}
The preliminary results of the luminosity measurement in a broad energy range are presented. 
The analysis is based on the integrated luminosity about $60 pb^{-1}$. 
For low energy range (smaller 320 MeV) the luminosity was determined using two processes: 
$e^+e^- \to e^+e^-, \mu^+\mu^-$. As for higher energies up to 2 GeV the luminosity determination was 
based on study processes $e^+e^- \to e^+e^-, \gamma\gamma$. As a result we had possibility to arrange 
cross-check and better estimate the systematics errors.
\subsubsection{Energy scan}
The precise determination of the luminosity is the key ingredient in many experiments 
which study the hadronic cross sections at $e^+e^-$ colliders. As a rule, the systematic 
error of the luminosity determination represents one of the largest sources of uncertainty 
which can cause significant reduction of the accuracy of the hadronic cross sections 
normalized to luminosity. Therefore it is very important to have several well known QED 
processes, for example,  $e^+e^- \to e^+e^-, \mu^+\mu^-, \gamma\gamma$ to determine 
the luminosity and to have cross check possibility.

The process $e^+e^- \to \gamma\gamma$ has essential advantages for luminosity~\cite{2g}.
It is free of difficulties 
related to both radiation of the final state particles and its Coulomb interaction. 
It is also of utmost importance that corresponding Feynman graphs do not contain 
photon propagators affected by the vacuum polarization effects. Events of this 
process have two collinear photons with similar energy deposition in calorimeters 
providing a clean signature for their selection. These reasons are the main motivation 
to explore this process as an independent tool for luminosity determination.     

The energy range from 1 to 2 GeV was scanned up and down with the energy step 50 MeV. 
At each energy point the integrated luminosity of about 500 nb$^{-1}$ was collected. 
During the scan down the energy points, at which data were collected, have been shifted 
to the previous one by 25 MeV. The energy of the 
beams has been monitored ($\sim$ 0.5 MeV) by measuring the current in an additional 
dipole magnet which was connected on in series with dipole magnets of the main ring.  

Two type of the first level triggers ``CHARGED'' and ``NEUTRAL'' were used while 
data taking. Signals from the drift and Z chambers
start a special processor ``TRACKFINDER'' (TF). ``CLUSTERFINDER'' (CF) was started 
by signals coming from calorimeters.  
A positive decision of any processor produces a command to write 
a current event on the online storage (space $\sim$ 2 TB).

\subsubsection {Luminosity determination}
At the first step the collinear events were selected 
according to the next criteria: ``CHARGED'' trigger produced positive decision; 
at least two tracks were reconstructed in DC; total charge must be equal to zero; distance 
of the both tracks from the beam axis is less than 0.5 cm; distance of the both tracks 
along beam axis from the interaction point does not exceed 10 cm; acollinearity angles 
between two tracks in the scattering plane (contains the beam axis), 
$ |\Delta \Theta| = |\Theta_{1} - (\pi - \Theta_{2})| \leq$0.25 rad; 
acollinearity angles between two tracks in the azimuthal plane 
(perpendicular to the beam axis), 
$ |\Delta \Phi| = |\pi - |\Phi_{1} -  \Phi_{2}|| \leq $0.15 rad; average polar 
angle of two tracks $ [\Theta_{1} + (\pi - \Theta_{2})]/2 $ 
should be between 1 and ($\pi-1$) rad.

The energy deposition of these events in calorimeters was 
used to separate $e^+e^-$ particles and determine their number. 
Thus, the integrated luminosity can be determined by the selected Bhabha events 
by the standard way.

\begin{figure}
\centering
\includegraphics[width=0.8\linewidth]{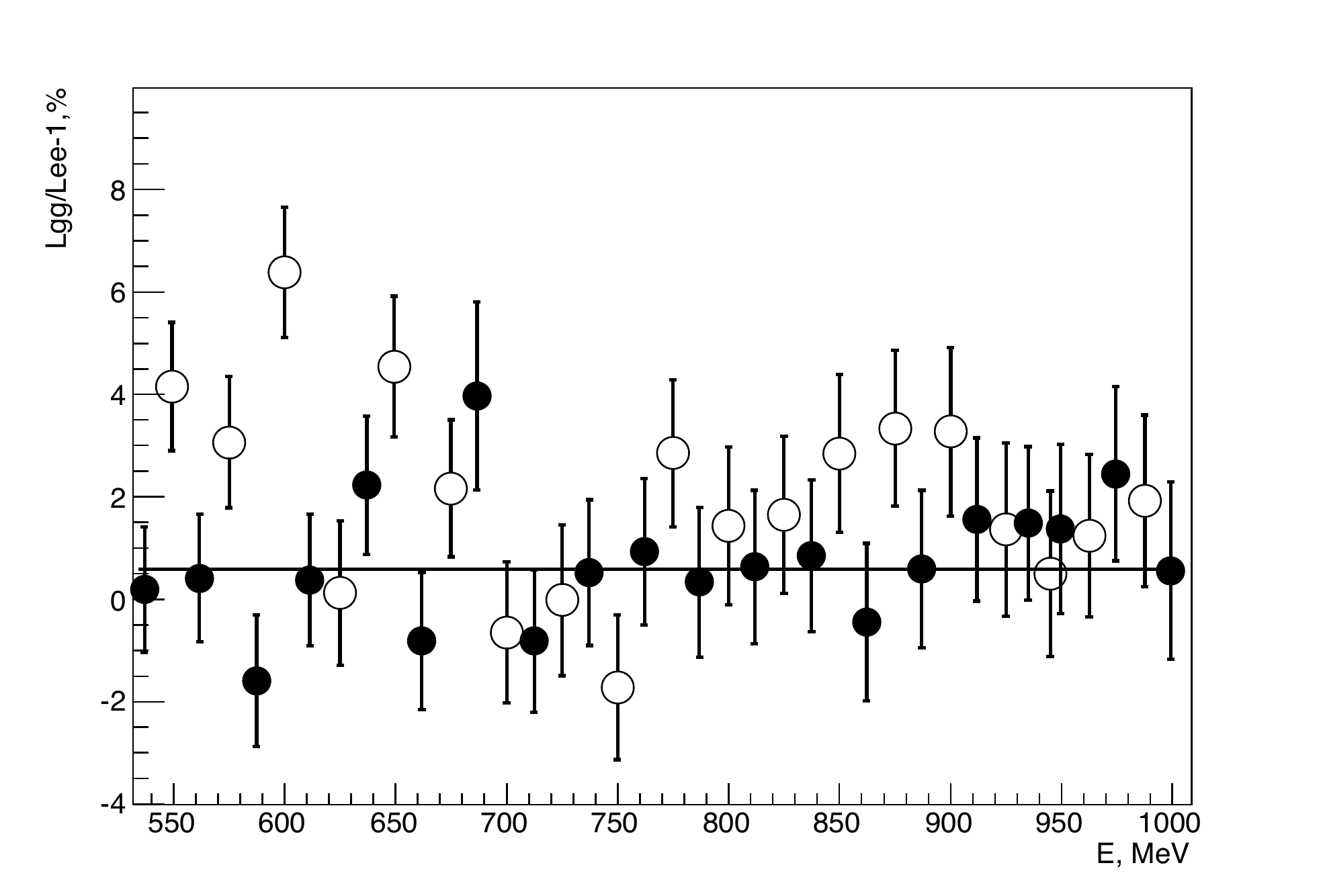}
\caption{The ratio of the relative difference of the luminosities vs beam energy. 
Empty circles - scan up, full circles - scan down.}
\label{lum-en-scan}
\end{figure} 

Events of the process $e^+e^- \to \gamma\gamma$ were also used to determine 
the integrated luminosity. To do that neutral collinear events were selected   
according to the following criteria: back-to-back clusters in the barrel 
calorimeters; no tracks in DC coming 
from the interaction point of the beams and no hits are in Z-chamber sectors 
associated with clusters. It is obvious the luminosity, 
determined by $\gamma\gamma$ events, has absolutely different systematic.
The ratio of the relative 
difference of the luminosities is presented in Fig.~\ref{lum-en-scan}, where only statistical 
errors are shown. The empty circles correspond to the scan up, whereas full circles - 
scan down. The horizontal line is a fit for this ratio for the scan down.
In this case the relative difference between luminosities is in average
smaller than 1\%. However, at the beginning of the run the difference was
about $\sim$5\% and explained by hardware problems and the quality of 
inter-calibration of the detector subsystems. Collecting all the main sources which
contribute to systematic error we estimate its accuracy as $\sim$2\%.
\begin{figure}
\centering
\includegraphics[width=0.8\linewidth]{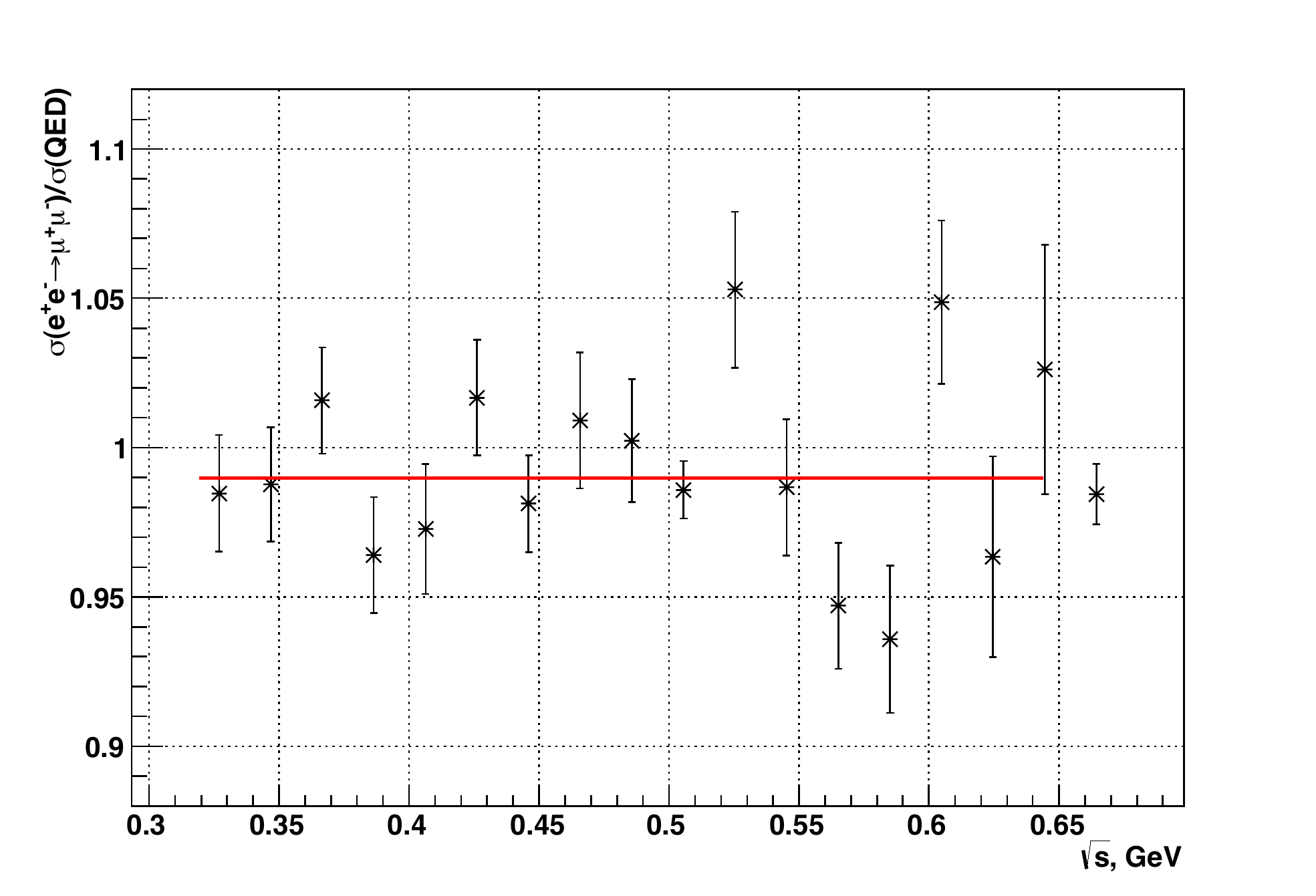}
\caption{The ratio of the relative difference of the luminosities vs beam energy.}
\label{mom-sep}
\end{figure} 
At energies 0.36 - 0.65 GeV momentum resolution of the DC enough to separate $e/\mu/\pi$ 
events. As it is shown in Fig.~\ref{mom-sep} the relative difference is about 
$(-1 \pm 0.5)$\%. In future runs we plan to have statistical error at the level 0.1\% - 0.2\% 
and directly check the Monte Carlo Generator Photon 
Jets~\cite{mcgpj} with radiative corrections with the similar accuracy.  

This work is supported in part by the Russian Education and Science
Ministry, by FEDERAL TARGET PROGRAM "Scientific and scientific-pedagogical personnel of 
innovative Russia in 2009-2013", by agreement 14.B37.21.07777,   by the Russian Fund for
Basic Research grant
RFBR 10-02-00695-a,
RFBR 10-02-00253-a,
RFBR 11-02-00328-a,
RFBR 11-02-00112-a,
RFBR 12-02-31501-a,
RFBR 12-02-31499-a,
RFBR 12-02-31498-a,
and RFBR 12-02-01032-a.

\newpage

\subsection{A combined estimate of the KLOE08, KLOE10 and KLOE12 ISR measurements}
\addtocontents{toc}{\hspace{2cm}{\sl S.~E. M\"uller}\par}
\label{sec:Mueller}
\vspace{5mm}

S.~E.~M\"uller

\vspace{5mm}

\noindent
Institute of Radiation Physics, Helmholtz-Zentrum Dresden-Rossendorf, Germany\\

\vspace{3mm}
The KLOE experiment at the Frascati $\phi$-factory DA$\Phi$NE has published 4 data sets for the cross section of the process $e^+e^- \to \pi^+\pi^-$ below 1 GeV~\cite{Aloisio:2004bu,Ambrosino:2008aa,Ambrosino:2010bv,Babusci:2012rp}. Since the use of the KLOE05~\cite{Aloisio:2004bu} data set in high precision applications is disfavoured due to a possible inconsistency in the trigger evaluation, the KLOE08~\cite{Ambrosino:2008aa} data set is considered to supersede the KLOE05 data. In the following, a combined estimate is constructed from the 3 remaining data sets KLOE08~\cite{Ambrosino:2008aa}, KLOE10~\cite{Ambrosino:2010bv} and KLOE12~\cite{Babusci:2012rp} and their covariance matrices~\cite{Czyz:2013zga}.

Based on the method of the {\it best linear unbiased estimator} (BLUE)~\cite{Lyons:1988rp,Valassi:2003mu}, from the 195 data points in total~\cite{KLOEurl} a best estimate is found for the 85 observables (corresponding to the 85 bins of $0.01$ GeV$^2$ binwidth between 0.10 and 0.95 GeV$^2$). Due to a possible bias in the presence of normalization errors~\cite{dagostini}, special care needs to be devoted to the construction of the $(195 \times 195)$ covariance matrix for the 195 input data points. A way out is to split the covariance matrix in 2 parts: A statistical covariance matrix that is used to construct the BLUE values, and a systematic covariance matrix that contains the normalization errors and is propagated properly a posteriori to the $(85 \times 85)$ covariance matrix of the best estimates.

Using the best linear unbiased estimates and their covariance matrix obtained in this way, an evaluation of the dispersion integral for $\Delta a_\mu^{\mathrm{\pi\pi}}$ yields
\begin{eqnarray}
\Delta a_\mu^{\pi\pi}[0.10-0.95 \mathrm{GeV}^2] & = &(488.6\pm5.7)\times 10^{-10}\\
\Delta a_\mu^{\pi\pi}[0.35-0.85 \mathrm{GeV}^2] & = &(378.9\pm2.8)\times 10^{-10}
\end{eqnarray}

This can be compared to the results obtained using a combination of the KLOE08 and KLOE10 only~\cite{KN225}, in which one obtains
\begin{eqnarray}
\Delta a_\mu^{\pi\pi}[0.10-0.95 \mathrm{GeV}^2] & = & (488.6\pm6.0)\times 10^{-10}\\
\Delta a_\mu^{\pi\pi}[0.35-0.85 \mathrm{GeV}^2] & = & (379.1\pm2.9)\times 10^{-10}
\end{eqnarray}

As can be seen, the addition of the KLOE12 data does not change too much. This can be explained with the fact that the only additional independent contribution comes from the $\mu\mu\gamma$-spectrum used in the KLOE12 analyses. This spectrum is statistically limited compared to the $\pi\pi\gamma$-spectra, and therefore does not influence much the BLUE values.

As a final step, one can update the correction for vacuum polarization used in the published values of the KLOE08 and KLOE10 data sets from the {\tt alphaqED03} to the {\tt alphaQED12} software package~\cite{Jegerlehner_alpha}, because the  {\tt alphaqED03} package shows some discrepancy when compared to more recent evaluations for the vacuum polarization correction~\cite{Actis:2010ggg}. Then constructing the BLUE values and evaluating $\Delta a_\mu^{\mathrm{\pi\pi}}$ gives the values
\begin{eqnarray}
\Delta a_\mu^{\pi\pi}[0.10-0.95 \mathrm{GeV}^2] & = &(487.8\pm5.7)\times 10^{-10}\\
\Delta a_\mu^{\pi\pi}[0.35-0.85 \mathrm{GeV}^2] & = &(378.1\pm2.8)\times 10^{-10}\, ,
\end{eqnarray}
a change by $0.7 - 0.8 \times 10^{-10}$, consistent with what was found in~\cite{KN225}. 
\\
\vspace{0.5mm}
The author wishes to thank the organizers of the ``PHIPSI13 - International Workshop on e+e- collisions from Phi to Psi 2013'' for support.

\newpage
\subsection{Hadronic light-by-light scattering in the muon $g-2$:
current status, open problems and impact of form factor measurements}
\addtocontents{toc}{\hspace{2cm}{\sl A.~Nyffeler}\par}
\label{sec:Nyff}
\vspace{5mm}

A.~Nyffeler

\vspace{5mm}

\noindent
Regional Centre for Accelerator-based Particle Physics, 
Harish-Chandra Research Institute, Chhatnag Road, Jhusi, Allahabad -
211019, India; nyffeler@hri.res.in \\ 
\vspace{3mm}

The hadronic light-by-light (HLbL) scattering contribution to the muon
$g-2$ involves the Green function of four electromagnetic currents,
connected to off-shell photons, see Ref.~\cite{JN_09_N_09} for details
and early references. In contrast to the hadronic vacuum polarization
(HVP) contribution, it cannot be directly related to experimental data
and therefore various hadronic models have been used to estimate
HLbL. But the dependence on several momenta leads to a mixing of
long and short distances and makes it difficult to avoid a double
counting of quark-gluon and hadronic contributions. In
Ref.~\cite{deRafael_94} a classification of the different
contributions to HLbL based on the chiral and large-$N_c$ counting was
proposed. At leading chiral order, but subleading in $N_c$, appears a
charged pion-loop. Leading in $N_c$, but subleading in the chiral
counting, are the exchanges of the light pseudoscalars $\pi^0, \eta,
\eta^\prime$. Also leading in $N_c$, but more suppressed in the chiral
counting, are the exchanges of other resonance states (axial-vectors,
scalars) and a quark-loop, representing the short-distance complement
of the low-energy hadronic models. In general, all the interactions of
the hadrons or quarks with the photons are dressed by some form
factors, e.g.\ via $\rho-\gamma$ mixing.

A selection of estimates for HLbL is presented in
Table~\ref{tab:summary}. Note that only Refs.~\cite{BPP,HKS} are full
calculations, using, as much as possible, one model for all the
contributions (ENJL in \cite{BPP}, HLS in \cite{HKS}).  The
compilations~\cite{HLbL_2007,PdeRV_09,JN_09_N_09} are based on these full
calculations, with revised or newly calculated values for some of the
contributions. More estimates, mostly for the pseudoscalar
contribution, can be found in Ref.~\cite{other_recent_papers}. While
most evaluations agree at the level of 15\%, if one takes the extreme
values, there is a spread of $a_\mu^{\rm HLbL; PS} = (59 - 107) \times
10^{-11}$.

\begin{table}[h]
  \centering
  \begin{tabular}{|c|c|c|c|c|c|c|}
    \hline 
    $\pi,K$-loops & $\pi^0,\eta,\eta^\prime$ & axial-vectors & scalars & 
    quark-loop & Total & Reference \\   
    \hline 
    $-19(13)$ & 85(13) & 2.5(1.0) & $-6.8(2.0)$ & 21(3) & 83(32) &
    \cite{BPP} \\  
    $-4.5(8.1)$ & 82.7(6.4) & 1.7(1.7) & - & 9.7(11.1) & 89.6(15.4) & 
    \cite{HKS} \\ 
    - & 83(12) & - & - & - & 80(40) & \cite{KN_PRD_02} \\ 
    0(10) & 114(10) & 22(5) & - & 0 & 136(25) & \cite{MV_04} \\ 
    - & - & - & - & - & 110(40) & \cite{HLbL_2007} \\ 
    $-19(19)$ & 114(13) & 15(10) & $-7(7)$ & 2.3 [c-quark] & 105(26) & 
    \cite{PdeRV_09} \\ 
    $-19(13)$ & 99(16) & 22(5) & $-7(2)$ & 21(3) & 116(40) &
    \cite{JN_09_N_09} \\
    - & 81(2) & - & - & 107(2) & 188(4) & \cite{FGW} \\
    $-(11-71)$ & - & - & - & - & - & \cite{EPR-M} \\ 
    $-20(5)$   & - & - & - & - & - & \cite{BZ-A} \\ 
    $-45$ & $+\infty$ & - & - & $60$ & - & undressed  \\ 
    \hline 
  \end{tabular}
  \caption{Summary of selected estimates for the different
    contributions to $a_\mu^{\rm HLbL} \times 10^{11}$. For
    comparison, the last line shows some results when no form factors 
    are used.}  
  \label{tab:summary}
\end{table}

Until 2010, a consensus had been reached about the central value
$a_\mu^{\rm HLbL} = 110 \times 10^{-11}$, but there was a discussion
on how to estimate the error, more progressively, $\pm 26 \times
10^{-11}$, in Ref.~\cite{PdeRV_09} and more conservatively, $\pm 40
\times 10^{-11}$, in Ref.~\cite{JN_09_N_09}. In view of the precision
goal of future $g-2$ experiments at Fermilab and
J-PARC~\cite{Fermilab_J-PARC} with $\delta a_\mu = 16 \times 10^{-11}$
and the continued progress in improving the error in HVP, the HLbL
contribution might soon be the main uncertainty in the theory
prediction, if it cannot be brought under better control.

In the last few years, several works have appeared which yield much
larger (absolute) values for some of the contributions, see
Table~\ref{tab:summary}. In Ref.~\cite{FGW} the quark-loop was studied
using a Dyson-Schwinger equation approach. In contrast to
Refs.~\cite{BPP,HKS}, no damping compared to the bare constituent
quark-loop result was seen, when a dressing was included. Note that
this calculation of the quark-loop is not yet complete. The large size
of the quark-loop contribution in Ref.~\cite{FGW} was questioned in
the papers~\cite{quark-loop}, using different quark-models and
approaches.  The pion-loop contribution was analyzed in
Ref.~\cite{EPR-M}. The authors stressed the importance of the
pion-polarizability effect and the role of the axial-vector resonance
$a_1$, which are not included in the models used in
Refs.~\cite{BPP,HKS}. Depending on the value of the
pion-polarizability and the model for the $a_1$ used, a large
variation was seen. The issue was taken up in Ref.~\cite{BZ-A} where
different models for the pion-loop were studied. The inclusion of the
$a_1$ was attempted, but no finite result for $g-2$ could be
achieved. With a cutoff of $1~\mbox{GeV}$, a result close to the
earlier estimate in Ref.~\cite{BPP} was obtained.

Concerning the future, maybe lattice QCD will provide a reliable
calculation of HLbL at some point (see Ref.~\cite{HLbL_Lattice} for
some promising recent results), but in the meantime only a close
collaboration between theory and experiment can lead to a better
controlled estimate for HLbL. From the theory side, the hadronic
models can be improved by short-distance constraints from perturbative
QCD to have a better matching at high momenta. Also the issue about
whether the dressing of the bare constituent quark-loop leads to a
suppression or an enhancement needs to be studied further.

From the experimental side, the information on various processes
(decays, form factors, cross-sections) of hadrons interacting with
photons at low and intermediate momenta, $|q|~\leq~2~\mbox{GeV}$, can
help to constrain the models. Important experiments which should be
pursued include more precise measurements of the (transition) form
factors of light pseudoscalars with possibly two off-shell photons in
the process $e^+ e^- \to e^+ e^- P \ (P= \pi^0, \eta, \eta^\prime)$
and the two-photon decay widths of these mesons. This could further
reduce the error of the dominant pseudoscalar exchange
contribution~\cite{KLOE-2_impact}. Concerning the pion-loop
contribution, in addition to studying $\gamma\gamma \to \pi\pi$,
measurements of the pion-polarizability in various processes, e.g.\ in
radiative pion decay $\pi^+ \to e^+ \nu_e \gamma$, in radiative pion
photoproduction $\gamma p \to \gamma^\prime \pi^+ n$ or with the
hadronic Primakoff effect $\pi A \to \pi^\prime \gamma A$ or $\gamma A
\to \pi^+ \pi^- A$ (with some nucleus $A$), should help to improve the
models~\cite{EPR-M}. For the development of models with $a_1$ and
estimates of the sizable axial-vector contribution, information about
the decays $a_1 \to \rho\pi, \pi\gamma$ would be useful as well. To
extract the needed quantities from experiment will also require the
development of dedicated Monte-Carlo programs for the relevant
processes.

If the recent results for the quark-loop and pion-loop are taken at
face value, one obtains the range $a_\mu^{\rm HLbL} = (64 - 202)
\times 10^{-11}$. While the new approaches raise some important issues
and point to potential shortcomings in the previously used models,
these estimates are also still preliminary and further studies are
needed. The estimate $a_\mu^{\rm HLbL} = (116 \pm 40) \times 10^{-11}$
from Ref.~\cite{JN_09_N_09} therefore still seems to give a fair
description of the current situation.

The author wishes to thank the organizers of the Working Group on
Radiative Corrections and Generators for Low Energy Hadronic Cross
Section and Luminosity, as well as the Heinrich-Greinacher-Stiftung,
Physics Institute, University of Bern, Switzerland, for support. He is
also grateful for the kind hospitality at the Albert Einstein Center
for Fundamental Physics, Institute for Theoretical Physics, University
of Bern, Switzerland.

\newpage

\subsection{Are $\tau^- \to \pi^- \ell^+ \ell^- \nu_\tau$ decays within discovery reach in near future?}
\addtocontents{toc}{\hspace{2cm}{\sl P.~Roig}\par}
\label{sec:Roig}
\vspace{5mm}

P.~Roig

\vspace{5mm}

\noindent
Instituto de F\'{\i}sica, Universidad Nacional Aut\'onoma de M\'exico, \\AP 20-364, M\'exico D.F. 01000, M\'exico\\
\vspace{3mm}

These decays \cite{Guevara:2013wwa} remain yet undetected, although being the crossed channels of the $\pi^-\to\ell^-e^+e^-\nu_l$ decays, measured some time ago. 
These processes probe the $W^\star-\gamma^\star-\pi$ vertex, with both gauge bosons off-shell and complement the $\tau^-\to\pi^-\gamma\nu_\tau$ and $\pi^-\to\ell^-\gamma\nu_l$ 
decays, which are sensitive to the $W^\star-\gamma-\pi$ vertex. Its knowledge in the full kinematical regime is important for testing QCD predictions, computing the radiative 
corrections to the non-photon processes and in the evaluation of the hadronic light-by-light contribution to the anomalous magnetic moment of the muon (through the vector part 
of the weak current). Hadronic tau decays are a powerful tool to extrapolate between the known chiral and asymptotic regimes.

The examined decays span different energy regions according to the virtualities of the exchanged $W$ and $\gamma$. For low momentum transfers, Chiral Perturbation Theory is 
the effective field theory of QCD and determines the behavior of the form factors in the so-called chiral limit. At larger energies the associated expansion breaks down but 
$1/N_C$ has proved to be an adequate alternative expansion parameter to enlarge the domain of applicability up to $M_\tau$. Resonance Chiral Theory~\cite{RChT} is a 
convenient realization of these ideas for the light-flavored mesons that we have employed. It is built upon the known chiral symmetry breaking and discrete symmetries of QCD 
and unitary symmetry for the resonances without any ad-hoc dynamical assumption. Although it yields an infinite number of states at leading order, the 
$\tau^-\to\pi^-\nu_\tau\ell^+\ell^-$ decays damp completely the contribution of excited resonances erasing any dependence on the realization of the spectrum made by an 
infinite tower of resonances. We have also included the leading next-order correction, given by the energy-dependent off-shell widths~\cite{GomezDumm:2000fz}.

These decays are generated by making the photon in the one-pion radiative tau decays virtual, then it converts into a lepton pair. Consequently, analogous 
contributions are obtained: inner bremsstrahlung off the tau, off the pion or from the local $W\gamma\pi$ vertex and model dependent parts with the hadronization of the 
(axial-)vector current. The different interference terms are non-vanishing and sizable, in general. The hadronic form factors depend on the photon off-shellness and on the 
product of the photon and pion momenta. A vanishing diagram for on-shell photon \cite{Guo:2010dv} contributes in this case, being proportional to the isovector part of the 
di-pion vector form factor, for which a dispersive representation \cite{Dumm:2013zh} describing successfully data was adopted.

Hadronic form factors must satisfy QCD short-distance behavior, which implies relations among the Lagrangian couplings that are in agreement with previous results 
\cite{RChT, Guo:2010dv, Dumm:2013zh, Literature} and we can predict the phenomenology of these decays. A conservative variation of one fifth was allowed on these relations in 
order to estimate the error of the high-energy constraints \cite{Roig:2012zj}. The branching ratio of the $\tau^-\to\pi^-\nu_\tau\ell^+\ell^-$ decays ($\ell=e,\,\mu$) are 
$\left(1.7^{+1.1}_{-0.3}\right)\cdot 10^{-5}$ ($e$) and $\left[3\cdot10^{-7},1\cdot10^{-5}\right]$ ($\mu$). According to these results the $\ell=e$ decays should be 
discovered soon at Belle-II or at the Italian or Russian $\tau-c$ factories. On the contrary, this is not guaranteed for the $\ell=\mu$ decays which would deserve a dedicated 
search in any case at near-future facilities.

Structure-dependent effects amount to $\sim15\%$ and $\sim92\%$ of the decay width for ($\ell=e,\,\mu$), respectively. The dominance of the model-dependent part in the 
di-muon case results in much larger quoted errors. In Ref.~\cite{Guevara:2013wwa} we analyze in detail the normalized di-lepton invariant mass distribution in both cases. 
The inner-bremsstrahlung contribution dominates in either case up to $\sim0.1$ GeV$^2$ where the axial-vector contribution overtakes it. This causes a change in the slope of 
the curve that can be easily measured even with very few events. In case a fine binning and enough statistics are achieved, the $\rho(770)$ contribution (through the $I=1$ 
pion vector form factor) will show up as a prominent peak.

These matrix elements will be coded in the new TAUOLA hadronic currents \cite{TAUOLA}. The present study is also relevant for better characterizing the associated background for lepton flavour violating searches \cite{Guo:2010ny} in the 
$\tau^-\to\mu^-\ell^+\ell^-$ process.
\\
\vspace{0.5mm}
I thank the WG organizers for their excellent job. This work has been partially funded by Conacyt and DGAPA. The support of project PAPIIT IN106913 is also acknowledged.


\newpage
\subsection{The tau decay in three pions in Resonance Chiral Theory. \\
Status of analysis}
\addtocontents{toc}{\hspace{2cm}{\sl O.~Shekhovtsova}\par}
\label{sec:Shek}
\vspace{5mm}
O. Shekhovtsova~\footnote{In collaboration with I. M. Nugent, T. Przedzi\'nski, 
P. Roig and Z. W\c{a}s}\\
\vspace{5mm}
\noindent Institute of Nuclear Physics, PAN,
        Krak\'ow, ul. Radzikowskiego 152, Poland
\vspace{3mm}

In our paper \cite{Shekhovtsova:2012ra} we described an upgrade of the Monte Carlo generator TAUOLA using 
the results of 
the Resonance Chiral Lagrangian ($R\chi L$) for the $\tau$ lepton decay into the most important two and three meson
 channels. 
The necessary theoretical concepts were collected, 
numerical tests of the implementations were completed and documented. Finally, we presented  strategy for fitting 
experimental data and calculated the systematic uncertainties associated with the experimental measurement. 
However, there was and remain until now,  an obvious limitation due to 
the fact that we are using one-dimensional projections of the invariant masses 
of a multi-dimensional distribution.
The first comparison \cite{Shekhovtsova:2013rb} of the $R\chi L$ results for the $\pi^-\pi^-\pi^+$ mode with the 
BaBar data 
\cite{Nugent:2013ij}, did not demonstrate a satisfactory agreement for the two pion invariant mass distributions.
With the recent availability of the unfolded distributions 
for all invariant masses constructed from observable decay products for this channel \cite{Nugent:2013ij},
we found ourselves in an excellent position to work on model improvement for the $\pi^-\pi^-\pi^+$ mode.  The choice of this channel is  motivated by its
relatively large branching ratio, availability of unfolded experimental distribution and already non-trivial 
dynamics of three-pion final state.  
In addition, this channel is important for Higgs spin-parity studies through the associated di-$\tau$ decays. 

The $R\chi L$ is devised for ordinary $q\bar{q}$ resonances. As
the  $\sigma$ meson is, predominantly, a tetraquark state, it cannot be included in the $R\chi L$ formalism. 
We have decided to incorporate the $\sigma$ meson as an extension of the phenomenological approach used by 
the {\tt CLEO} collaboration \cite{Asner:1999kj}. 
The fit values of the parameters used in our new model to BaBar data \cite{Nugent:2013ij} are collected in Table \ref{tab:fit}.  
The goodness of fit is quantified by $\chi^2/ndf = 6658/401$; that is eight times better than the 
previous result \cite{Shekhovtsova:2013rb}.

We also have estimated the effects of the electromagnetic interaction among the final-state pions.
The Coulomb interaction can be important near the production threshold. We use  
the far-field approximation;
the final-state pions  are treated as stable point-like objects and 
the three pion interaction is treated as a superposition of the two pion ones. 
 To estimate the Coulomb interaction in S-wave  
we can apply the results of Section 94 of  Ref.~\cite{Landau:101813} and consequently neglect P-wave Coulomb interaction among the final-state pions
as the precision studies of the P-wave two pion form factors does not require to include the Coulomb interaction. 
Taking into account the Coulomb interaction (when the $\sigma$ contribution is not included)
we obtain a $\chi^2/ndf = 33225/401$.
Therefore, the Coulomb interaction without the $\sigma$ contribution cannot describe the data in the low-energy region. 
\begin{table} 
\footnotesize
\begin{tabular}{|l|l|l|l|l|l|l|l|l|}
\hline
      & $M_\rho$   & $M_{\rho'}$& $\Gamma_{\rho'}$& $M_{a_1}$& $M_\sigma$& $\Gamma_\sigma$& $F$& $F_V$ \\
\hline
Min & 0.767 &1.35 & 0.30 & 0.99 & 0.400 & 0.400 & 0.088 & 0.11  \\
\hline
Max &  0.780 &1.50& 0.50 & 1.25 & 0.550 & 0.700 & 0.094 & 0.25  \\
\hline
Fit  & 0.771849 & 1.350000 & 0.448379 & 1.091865 & 0.487512 & 0.700000 & 0.091337 & 0.168652 \\
\hline
\end{tabular}
\begin{tabular}{|l|l|l|l|l|l|l|l|}
\hline
      & $F_{A}$ & $\beta_{\rho'}$ & $\alpha_\sigma$ & $\beta_\sigma$ & $\gamma_\sigma$ & $\delta_\sigma$ & $R_\sigma$ \\
\hline
Min   & 0.1 &  -0.37 & -10.  & -10.  &  -10.   & -10.      &  -10.     \\
\hline
Max   & 0.2 & -0.17 & 10.   & 10.  &  10.     &  10.     &  10.   \\
\hline
Fit   & 0.131425 & -0.318551 & -8.795938 & 9.763701 & 1.264263 & 0.656762 & 1.866913  \\
\hline
\end{tabular}
\vspace{-0.3cm}
\caption{Numerical ranges of the $R\chi L$ parameters
used to fit the BaBar data 
for three pion mode \cite{Nugent:2013ij}. }
\label{tab:fit}
\end{table}
\begin{figure}[h!]
\vspace{-0.2cm}
\centering
\includegraphics[scale=.260]{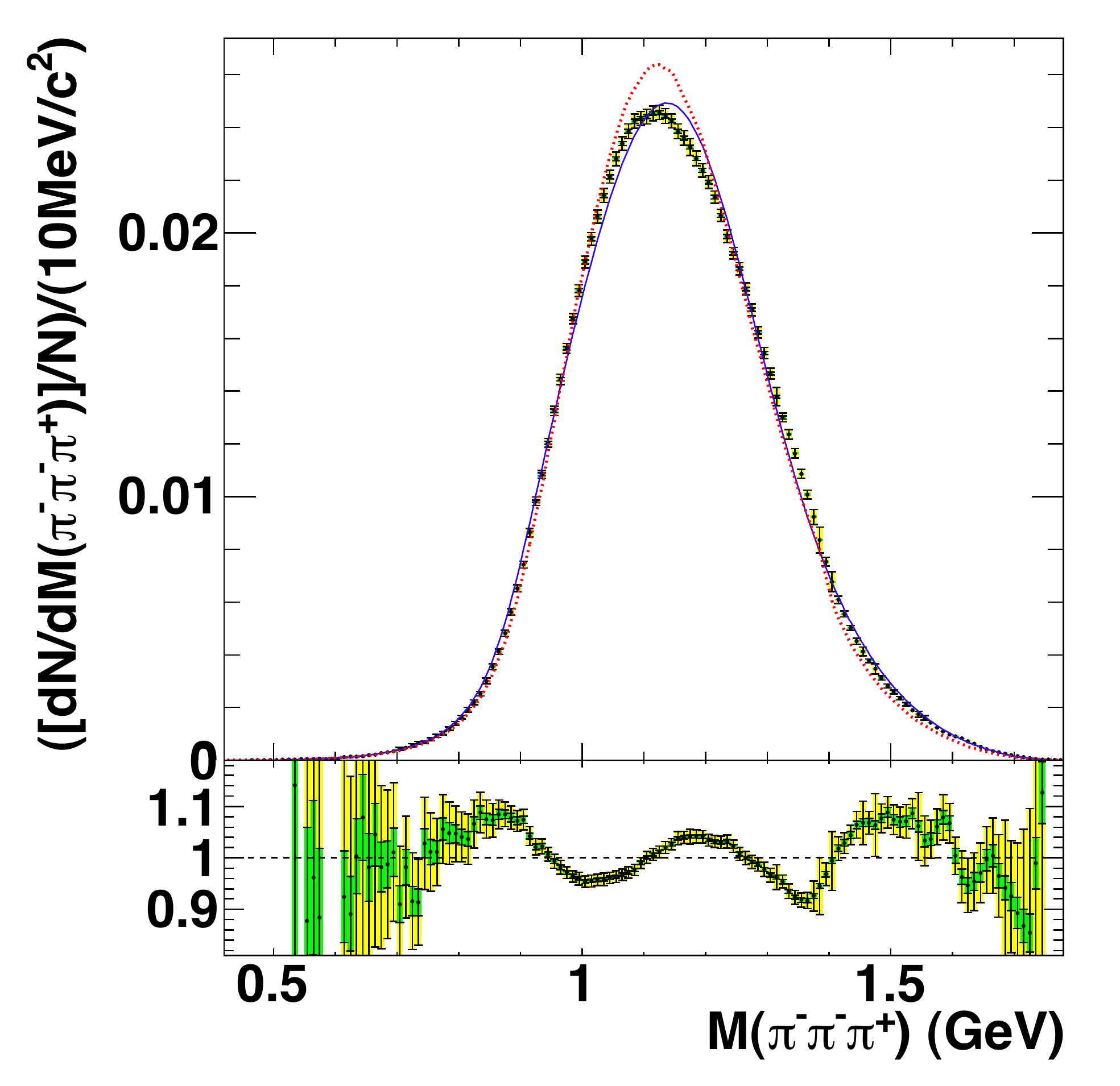}
\includegraphics[scale=.260]{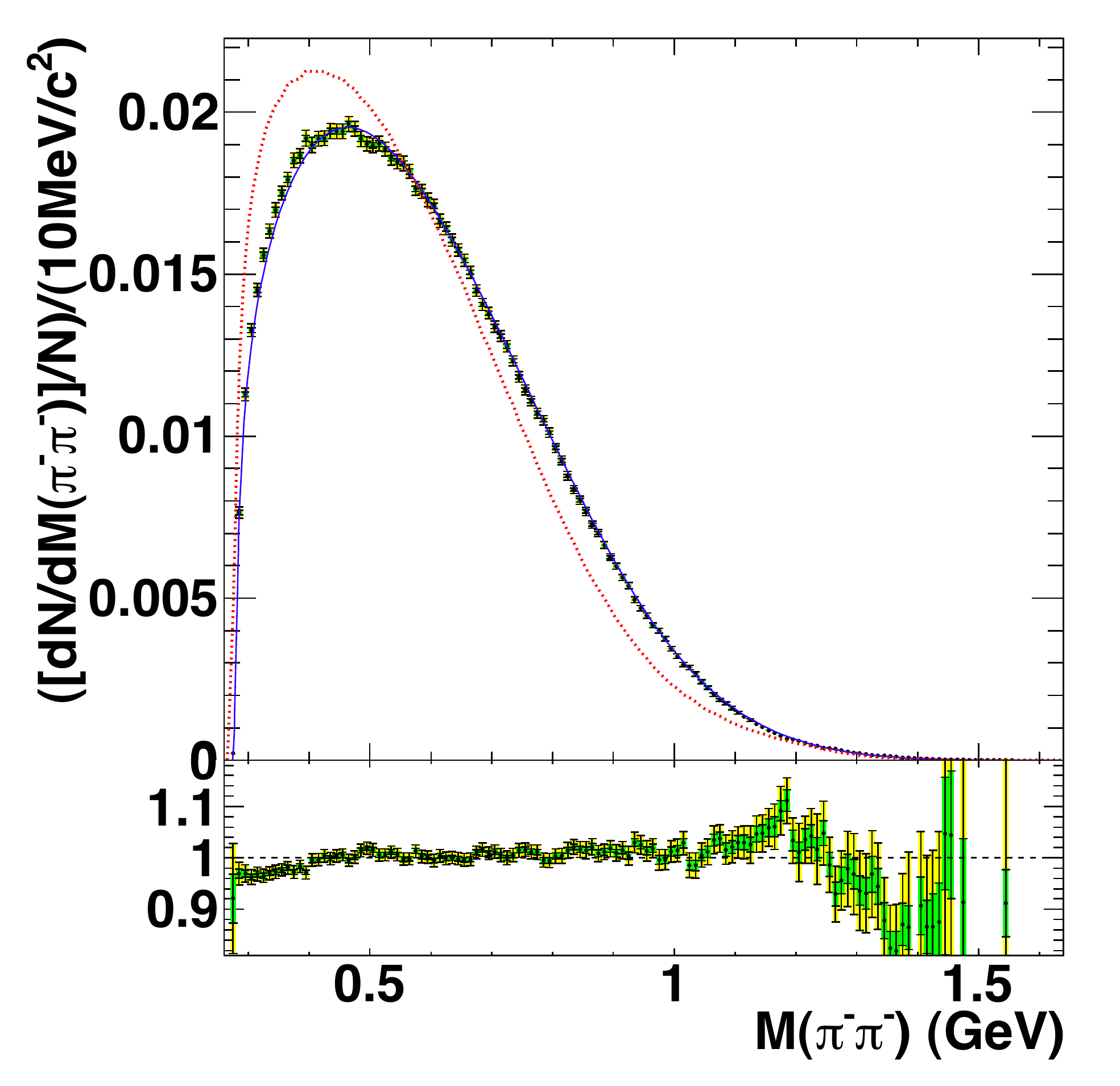}
\includegraphics[scale=.260]{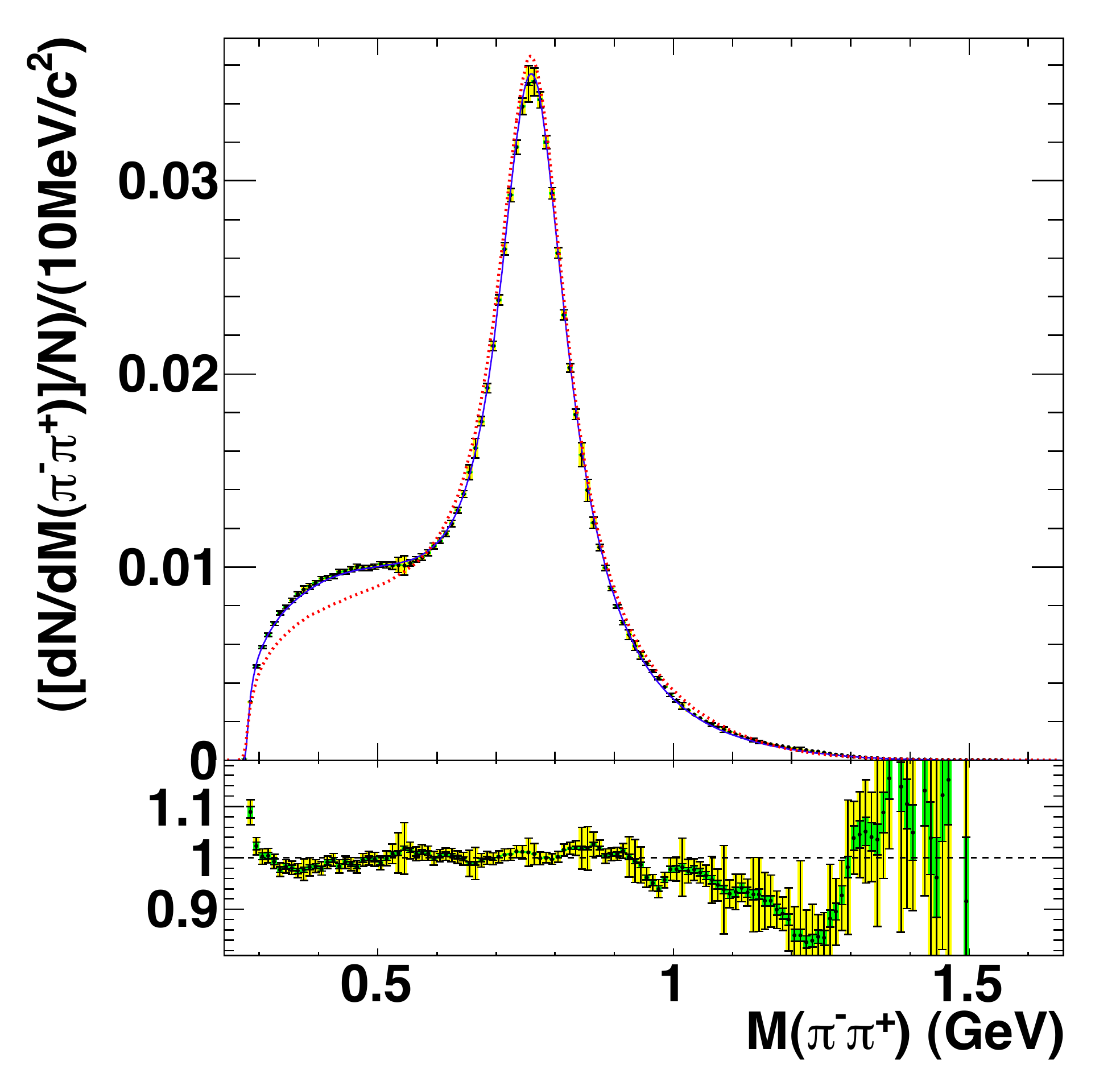}
\vspace{-0.6cm}
\caption{The differential decay width of the $\tau^- \to \pi^- \pi^-\pi^+\nu_\tau$ channel is plotted versus the  
invariant mass distribution of the three-pion system and two-pion pair systems. The BaBar measurements \cite{Nugent:2013ij} are represented 
by the data points, 
with the results from the $R\chi L$  current as described in the text (blue line) and the
 old tune from {\tt CLEO} from Refs.~\cite{Davidson:2010rw} (red-dashed line) overlaid.  
At the bottom of the figure, ratio of new $R\chi L$ prediction to the data is given.
\label{Fig:pipipi}}
\end{figure}
For the details of fitting procedure and  discussion on the numerical value of parameters, see \cite{ourpaper}.

The author acknowledges that the final step of this project is financed, in part, by Foundation of Polish Science, 
grant POMOST/2013-7/12. POMOST Programme is co-financed from European Union, Regional Development Fund.

\newpage
\subsection{How to bring the error of the VP contributions down and
  how the WG could contribute in this important task?}
\addtocontents{toc}{\hspace{2cm}{\sl T.~Teubner}\par}
\label{sec:Teub}
\vspace{5mm}

T.~Teubner

\vspace{5mm}

\noindent
Department of Mathematical Sciences, University of Liverpool,
Liverpool L69 3BX, U.K.\\
\vspace{3mm}

At the PhiPsi13 conference several talks have covered the latest
developments for $g\!-\!2$ of the muon in detail, and especially the
hadronic vacuum polarisation (VP) contributions which still have the
biggest error of all Standard Model contributions. Here we highlight
issues which will need to be resolved in order to reduce the
uncertainty of the VP contributions significantly as required by the
new $g\!-\!2$ experiments. 

\vspace{2mm}
\noindent
1.\ While more hadronic cross section measurements have increased the
accuracy of the prediction in general, data combinations in the
most important $\pi^+\pi^-$ channel show an increasing tension
between different data sets based on the energy scan and Radiative
Return methods. For example, HLMNT~\cite{Hagiwara:2011af} get (in units of
$10^{-10}$) $a_\mu^{\pi\pi, {\rm w/out\ RadRet}}= 498.7 \pm 3.3$ for
the two pion contribution from direct scan data only, whereas
$a_\mu^{\pi\pi, {\rm with\ RadRet}}= 504.2 \pm 3.0$ when the data from
Radiative Return are included, which is a significant shift of $+5.5$
units. This pull-up is mainly a consequence of the BaBar
data~\cite{Aubert:2009ad}, and the tension between the various sets
prohibits a further shrinking of the error as would have been expected
given the quality of the individual sets. The latest $\pi\pi(\gamma)$
analysis from KLOE~\cite{Babusci:2012rp_2} confirms the earlier
measurements and will not significantly alter the picture once fully
included, see also~\cite{SMueller_PhiPsi13}. With more data in this
channel expected from the Novosibirsk experiments, KLOE-2 and possibly BESIII,
the situation will hopefully improve, but at the same time a better
understanding why the existing data are only marginally consistent
would be highly desirable. The combined expertise within the Working
Group on data analysis, Monte Carlo simulations, radiative corrections
and Radiative Return may allow to gain the required insights.

\noindent
At PhiPsi13 Maurice Benayoun presented results for the hadronic
contributions to $g\!-\!2$, combining $e^+e^-$ and $\tau$ spectral
function data and using combined fits based on a Hidden Local Symmetry
(HLS) model, see \cite{MBenayoun_PhiPsi13,Benayoun:2012wc}. He quoted
significantly smaller hadronic contributions resulting in a
discrepancy of up to $\sim 5\,\sigma$, depending on the details of the
analysis. While this seems to be incompatible with the numbers from
HLMNT, it is worth to note that his number for the direct estimate,
i.e.\ without the HLS model fit and using only direct scan data, is
very close to the corresponding number from HLMNT given
above. However, performing the HLS fit (and including the $\tau$ data)
leads to downwards shift, $-4.3$ units, and even bigger when the
Radiative Return data from KLOE are included. (They are not using the
BaBar data as they are incompatible in their fit.) These three
different shifts taken together explain the large difference in the
quoted total numbers (and sigmas) between HLMNT and Benayoun {\it et
  al.} 

\vspace{2mm}
\noindent
2.\ In the $K^+K^-$ channel the latest data from
BaBar~\cite{Lees:2013gzt} lead to a significant shift upwards. These
data are, certainly close to the $\phi$ resonance, inconsistent with
earlier measurements from SND and CMD2. BaBar also obtains a different
mass in their resonance fit, which may hint at an explanation of this
puzzle. While $K^+K^-$ is a subleading channel, such details must be
clarified to achieve the best possible combination and the smallest
error for the $g\!-\!2$ prediction. 

\vspace{2mm}
\noindent
3.\ The sum of the many (subleading) channels at energies below/around
2 GeV is actually fairly consistent with the predictions from
perturbative QCD. Many more data are becoming available and the
Working Group should play a major role in pushing for the most
relevant analyses. Manpower (or the lack of it) is a problem across
the experiments, and any additional funding for PostDoctoral
researchers would be highly welcome.

\vspace{2mm}
\noindent
4.\ The hadronic VP contributions as predicted by various groups
contain additional errors due to uncertainties in the treatment of
radiative corrections applied to some of the data sets. For HLMNT this
error is about $2\cdot 10^{-10}$. Further efforts, combining
theoretical with experimental expertise, should be made to re-address
this problem. First studies at Liverpool indicate that this should
allow to reduce the additional error substantially.

\vspace{2mm}
\noindent
Tackling these issues will be important to decrease the error of
the SM prediction of $g\!-\!2$ by a factor of two, and the Working Group may
become instrumental for this aim. For the VP contributions, a first
step could be the creation of a `commented database' of the required
hadronic cross section, with clear additional information and
recommendations w.r.t.\ the reliability of data sets, their systematic
errors and the treatment of radiative corrections. Such a database
could be supported by the PDG group at the IPPP Durham and could form
a well-defined project for a future funding application of the Working
Group.

\newpage

\section{List of participants}

\begin{flushleft}
\begin{itemize}
\item D.~Babusci, Laboratori Nazionali di Frascati dell'INFN, {\tt danilo.babusci@lnf.infn.it} 
\item E. Baldin, Novosibirsk State University, {\tt e.m.baldin@inp.nsk.su }
\item J. Crnkovic, University of Washington, {\tt jcrnkovi@uw.edu}
\item H.~Czy\.z, University of Silesia, {\tt henryk.czyz@us.edu.pl }
\item A.~Denig, Johannes Gutenberg-Universit\"at Mainz, {\tt denig@kph.uni-mainz.de}
\item T.~Dimova, NSU/BINP, {\tt baiert@inp.nsk.su}
\item V.~Druzinin, NSU/BINP, {\tt druzhinin@inp.nsk.su} 
\item S.~Eidelman, Novosibirsk State University, {\tt eidelman@mail.cern.ch }
\item G.~Fedotovich, Budker Institute of Nuclear Physics, {\tt fedotovich@inp.nsk.su}
\item S.~Ivashyn, NSC/KIPT, {\tt s.ivashyn@gmail.com}
\item B.~Kubis, Bonn University, {\tt kubis@hiskp.uni-bonn.de}
\item A.~Kupsc, Uppsala University, {\tt Andrzej.Kupsc@physics.uu.se }
\item I.~Logashenko, Budker Institute of Nuclear Physics, {\tt ivan.logashenko@gmail.com}
\item X.~Mo, Insitute of High Energy Physics Beijing \& CAS, {\tt moxh@ihep.ac.cn}
\item S.~E.~M\"uller, Helmholtz-Zentrum Dresden-Rossendorf, {\tt stefan.mueller@hzdr.de}
\item A.~Nyffeler, Harish-Chandra Research Institute, {\tt nyffeler@hri.res.in}
\item S.~Raspopov, V. N. Karazin Kharkiv National University, {\tt razserg@mail.ru}
\item P.~Roig, Instituto de F\'isica UNAM, {\tt proig@ifae.es }
\item S.~Serednyakov, Budker Institute of Nuclear Physics, {\tt seredn@inp.nsk.su}
\item O.~Shekhovtsova, Institute of Nuclear Physics Cracow, {\tt olga.shekhovtsova@ifj.edu.pl }
\item B.~Shwartz, Budker Institute of Nuclear Physics, {\tt shwartz@inp.nsk.su}
\item E.~Solodov, Budker Institute of Nuclear Physics, {\tt E.P.Solodov@inp.nsk.su} 
\item T.~Teubner, University of Liverpool, {\tt thomas.teubner@liverpool.ac.uk }
\item G.~Venanzoni, Laboratori Nazionali di Frascati dell'INFN, {\tt Graziano.Venanzoni@lnf.infn.it }
\item P.~Wang, IHEP Beijing \& CAS, {\tt wangp@ihep.ac.cn}
\item C.~Z.~Yuan, IHEP Beijing, {\tt yuancz@ihep.ac.cn}
\end{itemize}
\end{flushleft}

\end{document}